\documentstyle[aps]{revtex}

\newcommand{\tr}{{\rm tr}}
\newcommand{\ri}{{\rm i}}
\newcommand{\re}{{\rm e}}
\newcommand{\bs}{\mbox{\boldmath{$\sigma$}}}
\newcommand{\ttau}{\tilde{\tau}}
\newcommand{\I}[2]{I^{(#2)}[#1]}

\begin{document}
\title{Slow Decoherence of Superpositions of Macroscopically Distinct States}
\author{Daniel Braun$^{(1)}$, Petr A.~Braun$^{(1,2)}$ and Fritz Haake$^{(1)}$}
\address{$^{(1)}$ FB7, Universit\"at--GHS Essen, 45\,117 Essen,
Germany\\
$^{(2)}$ Department of Theoretical Physics, Institute of Physics,
Saint-Petersburg University, Saint-Petersburg 198904, Russia}

\maketitle
\begin{abstract}
\index{abstract} Linear
superpositions of macroscopically distinct 
quantum states (sometimes also called Schr\"odinger cat states) are usually
{\em almost 
immediately} reduced to a statistical mixture if exposed to
the dephasing influence of a dissipative environment. 
Couplings to the environment with a certain symmetry can lead to {\em slow}
decoherence, however. We
give specific examples of slowly decohering Schr\"odinger cat states in a
realistic quantum optical system and discuss 
how they might be constructed experimentally.
\end{abstract}

\section{Introduction}
A Schr\"odinger cat state is a linear superposition of
two quantum states that differ on a
macroscopic scale. Whereas superpositions of quantum states are
commonplace in the microscopic world, a superposition of macroscopically
distinct states is practically never observed. 
The understanding of {\em why} this is so has evolved
considerably with the development of the quantum mechanics of dissipative
systems \cite{Feynman63,Weiss93}. Dissipative  systems are systems that
are coupled to  an environment (also called ``heat bath'') with a large
number of degrees of 
freedom. The coupling allows for an exchange of energy between system and
bath. Therefore the motion of the system will slow down till only thermal
fluctuations in equilibrium with the heat bath are left. This effect is also 
present in classical mechanics and takes place on a classical time scale, 
$T_{class}$. But the coupling to the environment also
very rapidly destroys quantum mechanical interference effects, and leads to
an effectively classical behavior. The
decoherence arises on a time scale $T_{dec}$ which is typically much shorter
than 
$T_{class}$. As a general rule, if  a
seperation $D$ in phase space can be
assigned to the states involved in the Schr\"odinger cat state, the two time
scales behave as 
$T_{dec}/T_{class}\sim 
(\hbar/D)^r$ where $r$ is some positive 
power depending on the details of the coupling and the environment
\cite{Zurek81,Caldeira83,WallsMilburn,Haake87,Zukowski93,Strunz97}. This 
separation of time scales has also been termed ``accelerated decoherence''. 
The validity of this picture
starts to be confirmed  
in experiments of Haroche {\em et al.}, who were able to turn on
the coupling to an environment in a controlled way and measure the
decay of the coherences \cite{Brune96,Haroche98}. \\

The basis in which the density matrix becomes diagonal depends on the
coupling to the environment, as was pointed out by Zurek \cite{Zurek81}.
Suppose that the coupling Hamiltonian contains a ``coupling agent''
$\hat{A}$, $\hat{H}_{int}=\hat{A}f(\{ {\bf x},{\bf p} \})$,
where $f(\{{\bf x},{\bf p}\})$ is some function of the environmental
coordinates. Accelerated decoherence arises in the basis formed by the
eigenstates of $\hat{A}$, the socalled pointer basis. Thus, after the
short time $T_{dec}$ the reduced density matrix of the system will be
diagonal in this basis, but may still contain off--diagonal matrix elements
(coherences) in another. 

It has recently been shown, however, that even in the pointer basis
accelerated decoherence can be absent if the coupling to the environment has
a certain symmetry
\cite{Zurek81,Lidar98,GerryHach,GarrawayKnight,FilhoVogel,Zollergang}.
Indeed, suppose that 
$\hat{A}|\psi_1\rangle=a|\psi_1\rangle$ and
$\hat{A}|\psi_2\rangle=a|\psi_2\rangle$, i.e.~$a$ is a degenerate eigenvalue
of $\hat{A}$, then the environment cannot
distinguish between states $|\psi_1\rangle$ and $|\psi_2\rangle$.
A Schr\"odinger cat state $|\psi\rangle = c_1|\psi\rangle +c_2|\psi\rangle$
is therefore 
stable against dephasing. It may 
loose its coherence on the time scale on which 
the symmetrical arrangement of the state is destroyed, which can be as large
as $T_{class}$.\\ 

In this paper we give specific examples of such  longlived Schr\"odinger
cat states in the context of superradiance with a well known, experimentally
verfied 
damping mechanism. We calculate the decoherence rates and
propose a way how the Schr\"odinger cat states might be realized
experimentally.  

\section{Superradiance and the Schr\"odinger Cat States}

The system that  we consider is well known from the phenomenon of
superradiance \cite{Haroche82}: A large number of identical two--level atoms
in a
cavity radiate collectively. We assign formally to each atom 
a spin--$\frac{1}{2}$ operator ${\bf S}^{(i)}$ whose $z$--components tells us
whether the atom is in its lower or upper state. If all atoms couple in the
same way to a resonant electromagnetic field mode in the cavity,
the system as
a whole can be described by a collective variable ${\bf J}=\sum_{i=1}^N{\bf
S}^{(i)}$, the so called Bloch vector ($N$ is the number of atoms). It can
be thought of formally as an angular momentum. The coupling of the atoms to the
resonant electromagnetic field mode (with
creation and annihilation operators $a^\dagger$ and $a$) be of the form
\begin{equation}
\label{Hint}
 H_{int}=\hbar g(J_+a+J_-a^\dagger)\,,
\end{equation}
with some coupling constant $g$. Dissipation ultimately arises by the
cavity's non--ideal mirrors, i.e. the field mode is itself damped and 
photons may leak out of the cavity at a rate $\kappa$. Under the
assumption of weak coupling ($g\sqrt{N}/\kappa\ll 1$), low temperatures
($k_BT\ll \hbar \omega$, the level seperation of the atoms), and the Markov
assumption ($t\gg 1/\kappa$) the master equation 
\begin{equation} \label{eq:rhotd}
\frac{d}{d\tau}\hat{\rho}(\tau)=\frac{1}{2j}([J_-,\hat{\rho}(\tau) J_+]+[J_-\hat{\rho}(\tau), J_+])
\end{equation}
for the reduced  density operator $\hat{\rho}$ was derived in
\cite{Bonifacio71.1}.  We have written the  
time $t$ dimensionless in units of the inverse classical damping
rate as $\tau=2jg^2t/\kappa$, i.e.~$\tau$ is already in units of
$T_{class}$. The 
classical limit is obtained by 
$j\to\infty$ and yields a spherical 
phase space in which the orientation of the classical Bloch vector can be
parametrized by the polar angles $\theta$ (reckoned against the $J_z$--axis)
and $\varphi$ (the azimuth of the equatorial projection reckoned against the
$J_x$--axis). The variables 
$p=J_z/j\equiv \cos \theta$ and $\varphi$ aquire the role of 
classical canonical momentum and coordinate \cite{thebook}. Since the
classical phase space 
contains $2j+1$ states, we may think of $\hbar$ as scaling like $1/j$.
The Bloch vector behaves classically like an overdamped pendulum,
$d\theta/d\tau=\sin\theta$ and $\varphi=const.$.
Many of the consequences of (\ref{eq:rhotd}) have been confirmed experimentally
\cite{Haroche76}.\\ 

The states that correspond as much as possible to classical states are the
so called angular momentum coherent states 
$|\gamma\rangle=|\theta,\varphi\rangle$ \cite{Glauber72,Arecchietal}.  They
correspond to a classical angular momentum pointing in the direction given
by the polar angles $\theta$ and $\varphi$, with the complex label $\gamma$
given by
$\gamma=\tan(\theta/2)e^{i\varphi}$. They have minimal uncertainty, $\Delta
p\Delta q\sim 1/j$. In terms of $|jm\rangle$ states one has the
expansion 
\begin{equation} \label{cs}
|\gamma\rangle=(1+\gamma\gamma^*)^{-j}\sum_{m=-j}^j\gamma^{j-m}\sqrt{2j\choose
j-m}|jm\rangle\,. 
\end{equation}
Coherent states may be more familiar from the harmonic oscillator, where
they are eigenstates of the annihilation operator. The compactness of
Hilbert space prevents the existence of exact eigenstates of $J_-$, but one
can show that the angular momentum coherent states are approximate
eigenstates of $J_-$ in 
the sense that the angle between $J_-|\gamma\rangle$ and $|\gamma\rangle$
is of the order of $1/j$. They therefore qualify as pointer states in the
limit $j\to\infty$.   

In the following we will consider the fate of a Schr\"odinger cat state 
\begin{equation} \label{cat}
|\psi\rangle={\cal N}(|\gamma_1\rangle+|\gamma_2\rangle)\,,
\end{equation}
where ${\cal N}$ is the appropriate normalization constant. Note that the
two components can indeed be macroscopically distinct if the number of
atoms in the superradiance experiment is large.
We will show that  the damping (\ref{eq:rhotd}) leads in
general to accelerated decoherence. Our central prediction is,
however, that 
{\em accelerated decoherence is absent for Schr\"odinger cat states with
$\gamma_1\gamma_2^*=1$}.  Such two states correspond to two
classical angular momenta arranged symmetrically with respect to the equator
$\theta=\pi/2$ (i.e.~$\theta_2=\pi-\theta_1$) on a great circle
$\varphi_1=\varphi_2=const.$. They decohere only on the classical time
scale, $T_{dec}\sim T_{class}$. \\ 

The initial
density matrix corresponding to the state (\ref{cat}) reads
$\hat{\rho}(0)=|{\cal N}|^2(|\gamma_1\rangle\langle
\gamma_1|+|\gamma_1\rangle\langle \gamma_2|+|\gamma_2\rangle\langle
\gamma_1|+|\gamma_2\rangle\langle \gamma_2|)$. Since the 
evolution equation of $\hat{\rho}(\tau)$ is linear it suffices to
concentrate on one off--diagonal part $\tilde{\rho}(\tau)$ evolving from 
$\tilde{\rho}(0)=|\gamma_1\rangle\langle \gamma_2|$. Its decay is
conveniently studied in terms of one of the norms
\begin{eqnarray} \label{norms}
N_1(\tau)&=&\tr\tilde{\rho}\tilde{\rho}^\dagger\mbox{, or }\\
N_2(\tau)&=&\sum_{m_1,m_2}|\tilde{\rho}_{m_1\,m_2}|\,,
\end{eqnarray}
where $\rho_{m_1\,m_2}=\langle jm_1|\tilde{\rho}|jm_2\rangle$. Both norms
are obviously zero if coherences in the $\gamma$ basis are absent. The time
on which they decay to zero sets the decoherence time scale $T_{dec}$. 

Our prediction of slow decoherence, $T_{dec}\sim T_{class}$ for
$\gamma_1\gamma_2^*=1$, is 
based on three analytical results:       
\begin{enumerate}
\item The initial time derivative of $N_1(\tau)$ reads 
\begin{eqnarray} 
\frac{dN_1(\tau)}{d\tau}\left.\right|_{\tau=0}=&-&2j\left(\sin^2\theta_1+\sin^2\theta_2-
2\cos(\varphi_2-\varphi_1)\sin\theta_1\sin\theta_2\right)\\\label{der} 
&-&\left((1+\cos\theta_1)^2(1+\cos\theta_2)^2\right)\nonumber\,.
\end{eqnarray}
Thus, for $\varphi_2=\varphi_1$ and $\sin\theta_1=\sin\theta_2$, the term
proportional to $j$ vanishes. This marks the absence of accelerated
decoherence. \\
In the following we will restrict ourselves to
$\varphi_1=\varphi_2=0$ and denote $\gamma_i=|\gamma_i|$ throughout the rest
of the paper.
\item In the particular case of the states $|\gamma_1\rangle=|jj\rangle$ and
$|\gamma_2\rangle =|j\,-j\rangle$
(corresponding to $\gamma_1=0$ and $\gamma_2=\infty$), the exact time
evolution $N_2(\tau)$ is given by $N_2(\tau)=e^{-\tau}$
for all times! The 
coherence decays on the classical time scale, even though the two states are
macroscopically as distinct as possible. 
 
\item For finite times $\tau$ with $j\tau\ll 1$, a semiclassical evaluation
of the norm $N_2(\tau)$ for $\varphi_1=\varphi_2=0$ leads to
\begin{equation} \label{31}
\frac{N_2(\tau)}{N_2(0)}=\exp\left(-2j\frac{(\gamma_1-\gamma_2)^2(1-\gamma_1\gamma_2)^2}{((1+\gamma_1^2)(1+\gamma_2^2))^2}\tau\right)(1+{\cal
O}(1/j))\,.
\end{equation}
This means accelerated decoherence as long as $\gamma_1\ne\gamma_2$
and $\gamma_1\gamma_2\ne 1$. If, however, $\gamma_1\gamma_2=1$ then the next
order in $1/j$ shows that  
\begin{eqnarray} \label{32}
\frac{N_2(\tau)}{N_2(0)} &=&
\exp\left(-\left(\frac{\gamma_1^2-1}{\gamma_1^2+1}\right)^2\tau-\frac{3\gamma_1^8-3\gamma_1^6+4\gamma_1^4-3\gamma_1^2+3}{2(\gamma_1^2+1)^4}
\tau^2\right)\\
&&\cdot(1+{\cal O}(1/j))\,.
\end{eqnarray}
The expression in the exponent is correct up to and including order
$(j\tau)^2$. {\em Obviously, accelerated 
decoherence is absent for $\gamma_1\gamma_2=1$}. Indeed, a
single coherent state $\gamma_1=\gamma_2=\gamma$ leads, in linear order, to
almost the same decay, 
\begin{equation} \label{33}
\frac{N_2(\tau)}{N_2(0)}=\exp\left(-\gamma^4\left(\frac{\gamma^2-1}{\gamma^2+1}\right)^2\tau\right)\,.
\end{equation} 
\end{enumerate}

It is clear that the symmetry of the sine
function under $\theta\to\pi-\theta$ underlies the longevity. As mentioned the
coherent states $|\gamma\rangle$ are approximate eigenstates of the coupling
agent 
$J_-$ and therefore qualify as pointer states. The symmetry of the sine
function means that the corresponding ``approximate eigenvalue'',
$j\sin\theta e^{-i\varphi}$, is doubly degenerate.  A linear combination
of two vectors of the pertaining subspace is stable against
decoherence. We have here the interesting situation that the deviation of
order $1/j$ from
exact degeneracy is small enough to compensate for the acceleration factor
$j$ of the decoherence rates. Similiar but more general statements about the
decoherence in the presence of degeneracy breaking coupling
agents can be found in \cite{Lidar98}.\\    

We now describe in some detail the derivation of the three mentioned results.
The first, eq.(\ref{der}) can be obtained straight forwardly by inserting the
equation 
of motion for $\rho(\tau)$ into
$\frac{dN_1(\tau)}{d\tau}=\tr(\frac{d\rho}{d\tau}\rho^\dagger+\rho\frac{d\rho^\dagger}{d\tau})$.

The second result is an exact solution of the master
equation (\ref{eq:rhotd}) \cite{Bonifacio71.1,Braun98}. To see this we write
(\ref{eq:rhotd}) in the 
$|jm\rangle$--basis and introduce mean and relative quantum numbers,
$m=\frac{m_1+m_2}{2}$ and $k=\frac{m_1-m_2}{2}$. The latter quantum number
is conserved and therefore only enters as a parameter,
$\rho_{m_1,m_2}=\rho_m(k,\tau)$.  
The solution can be written  as 
\begin{equation} \label{sol}
\rho_m(k,\tau)=\sum_{n=-j+|k|}^{j-|k|}D_{mn}(k,\tau)\rho_n(k,0)
\end{equation}
in terms of the dissipative propagator $D_{mn}(k,\tau)$. With the help of
the abbreviations $Q_{mn}=\frac{(j+n)!(j-m)!}{(j+m)!(j-n)!}$ and
$g_l=j(j+1)-l(l-1)$  one has an
exact Laplace integral representation for $D_{mn}(k,\tau)$,
\begin{equation} \label{dmn}
D_{mn}(k,\tau)=\frac{\sqrt{Q_{m-k,n-k}Q_{m+k,n+k}}}{2\pi \ri}\int_{b-\ri
\infty}^{b+\ri\infty} ds\,e^{\tau s/j}\prod_{l=m}^n\frac{1}{s+g_l-k^2}\,,
\end{equation}
where the real parameter $b$ should be chosen larger than the largest pole
in the denominator.  
Depending on $m$ and $n$, in general a large number of poles contribute to
the integral. But for the case at our interest ($m_1=j$, $m_2=-j$), we have
$m=0=n$, $k=j$ and therefore only one pole contributes. We 
immediately obtain $N_2(\tau)=\rho_0(j,\tau)=D_{00}(j,\tau)\rho_0(j,0)=
\exp(-\tau)$.

Starting point for the third result is the short time propagator 
\begin{equation} 
D_{mn}(k,\tau)=\frac{\sqrt{Q_{m-k,n-k}Q_{m+k,n+k}}}{(n-m)!}\left(\frac{\tau}{j}\right)^{n-m}
\re^{\left\{-\frac{\tau}{j}\left[j^2-\left(\frac{n+m-1}{2}\right)^2\right]
\right\}}\label{smalltau1}
\end{equation}
derived in \cite{Braun98}. It is valid up to times
$\frac{|m+n-1|}{j}\,(n-m)\tau\ll 1$. Since we are going to sum over $m$ and
$n$, both $m+n$ and $m-n$ can be of
order $j$ and one should therefore have $j\tau\ll 1$. The
expansion (\ref{cs}) shows that (at $\varphi=0$) all initial density matrix
elements are 
real and positiv. Since also the short time propagator is real and positiv,
the same is true for the density matrix elements to the later time $\tau$,
such that the norm $N_2$ simplifies to 
\begin{equation} \label{n1}
N_2(\tau)=\sum_{k,m}\rho_m(k,\tau)=\sum_{m,n,k}D_{mn}(k,\tau)\rho_n(k,0)\,.
\end{equation}   
Note that in general $N_2(0)\ne 1$, such that in order to study the decay of
the coherences one should look at the quantity $n(\tau)\equiv
N_2(\tau)/N_2(0)$. \\ 
The first step in calculating $n(\tau)$ is to evaluate the sum over the
final states $m$. 
This summation does not envolve the initial density matrix
at all, so we can define $S(n,k,\tau)\equiv \sum_m D_{mn}(k,\tau)$ and 
$N(\tau)=\sum_{n,k}S(n,k,\tau)\rho_n(k,0)$. The latter two summations will
eventually be done by Laplace's method, that is by an asymptotic expansion
in $1/j$ for $j\to\infty$. In order to remain consistent with the limitation
$j\tau\ll 1$ we will have to rescale time as 
\begin{equation} \label{resc}
\tau=\frac{\tilde{\tau}}{j}
\end{equation}
with $\ttau\ll 1$ kept fixed for $j\to\infty$. This is indeed reasonable
since we expect decoherence of an ordinary Schr\"odinger cat  state on a
time scale 
$1/j$, that is $N(\tau)\simeq N(0)\exp(-j\tau)$. If we kept $\tau$ fixed for
$j\to\infty$ we could only recover $N(\tau)=0$, whereas we can
study the decay if we keep $\ttau$ fixed.
The precise calculation of
$S(n,k,\tau)$ by summing over $m$ is  difficult. However,
since we shall be interested only in short times $(j\tau \ll 1)$ we can
reconstruct $S(n,k,\tau)$ from its initial time derivatives. 
Let us write the derivatives directly in reduced coordinates that become
continous in the limit $j\to\infty$, $\nu=n/j$ and 
$\eta=k/j$, as 
\begin{eqnarray}
\frac{\partial S(n,k,\tau=\frac{\ttau}{j})}{\partial
\ttau}\left.\right|_{\ttau=0}&=&w_1+\eta^2-1+(\nu-\frac{1}{2j})^2 \nonumber \\
\frac{\partial^2 S(n,k,\tau=\frac{\ttau}{j})}{\partial\ttau^2}\left.\right|_{\ttau=0}&=&\left(\frac{\partial S(n,k,\tau=\frac{\ttau}{j})}{\partial
\ttau}\right)^2_{\ttau=0}+w_1(w_2-w_1)\\
&&+2w_1\left(\frac{-\nu}{j}+\frac{3}{4j^2}\right)\nonumber
\\
w_1&=&\sqrt{\left(1-(\nu-\eta-\frac{1}{2j})^2\right)\left(1-(\nu+\eta-\frac{1}{2j})^2\right)}\nonumber
\\
w_2&=&\sqrt{\left(1-(\nu-\eta-\frac{3}{2j})^2\right)\left(1-(\nu+\eta-\frac{3}{2j})^2\right)}\,.\nonumber
\end{eqnarray}
For the
further derivation we will keep $S$ in the 
form of a polynomial,
$S(\nu,\eta,\ttau)=1+a(\nu,\eta,1/j)\ttau+b(\nu,\eta,1/j)\ttau^2+{\cal
O}(\ttau^3)$ with $a(\nu,\eta,1/j)=\partial
S/\partial\ttau\left.\right|_{\ttau=0}$ and
$b(\nu,\eta,1/j)=\frac{1}{2}\partial^2
S/\partial\ttau^2\left.\right|_{\ttau=0}$. Higher terms in $\ttau$ can be
included, but already the first two 
terms account rather well for the initial decay of the coherences.\\

We now transform the remaining two sums in $N_2(\tau)$ (eq.(\ref{n1})) into
integrals by Euler--Maclaurin summation and finally integrate via Laplace's
method. To this end we first write the coefficients $\rho_n(k,0)$ in
the expansion of  the initial density matrix as continuous functions of the
reduced coordinates $\eta$ and $\nu\equiv n/j$. The result can be cast in
the form
\begin{equation} \label{rho0}
\rho_n(k,0)=C\re^{jS_0(\nu,\eta)}\,,
\end{equation} 
 where the ``action'' $S_0$ is given by
\begin{eqnarray} \label{S0}
S_0(\nu,\eta)&=&(1-\nu)\ln(\gamma_1\gamma_2)+\eta\ln\frac{\gamma_2}{\gamma_1}-\frac{1}{2}\big(p(1-\nu-\eta)\\
&&+p(1+\nu+\eta)+p(1-\nu+\eta)+p(1+\nu-\eta)\big)\\
p(x)&=&x\ln x\,.
\end{eqnarray}
The action is correct up to lowest order $1/j$. Terms of order $1/j$ have
been absorbed in the prefactor $C$ which turns out to be independent of
$\nu,\eta$ 
and therefore will cancel in $n(\tau)$. We thus have to order $\ttau^2$
\begin{equation} \label{Nint}
N_2(\tau)=2j^2C\int d\eta\int d\nu\left(1+a(\nu,\eta,1/j)\ttau+b(\nu,\eta,1/j)\ttau^2\right)\exp(jS_0(\nu,\eta))\,.
\end{equation} 
The prefactor 2 stems from the fact that $k$ and $n$ can be simultaneously
integer or semi--integer, but cancels of course as well in $n(\tau)$. Since
the slow decoherence is a $1/j$ effect, a carefull asymptotic expansion of
the double integral in powers  of $1/j$ is in order. First of all, we 
expand the coefficients $a$ and $b$ as $a(\nu,\eta,1/j)=\sum_{k=0}^\infty
a_k(\nu,\eta)\frac{1}{j^k}$ and $b(\nu,\eta,1/j)=\sum_{k=0}^\infty
b_k(\nu,\eta)\frac{1}{j^k}$. The first few terms in the
expansion read ($w=\sqrt{\left(1-(\nu-\eta)^2\right)\left(1-(\nu+\eta)^2\right)}$)
\begin{eqnarray}
a_0&=&w+\eta^2-1+\nu^2\label{a0}\\
a_1&=&\nu\frac{1-\nu^2+\eta^2-w}{w}\label{a1}\\
b_0&=&\frac{1}{2}a_0^2\label{b0}\\
b_1&=&(a_0a_1-\nu(\nu^2+w-\eta^2-1))\label{b1}\\
b_2&=&\frac{1}{2w^2}\Big((2a_0a_2+a_1^2)\big((\eta^2-\nu^2)^2-2(\eta^2+\nu^2)+1
\big)-2\nonumber\\
&&+4\nu^6-2\eta^6-10\nu^4\eta^2+8\nu^2\eta^4-10\nu^4-8\nu^2\eta^2+2\eta^4+8\nu^2+2\eta^2
\Big)\nonumber\\
&&+\frac{1}{4w}\Big(3\eta^4+7\nu^4-10\nu^2\eta^2-10\nu^2-6\eta^2+3\Big)\label{b2}\,.
\end{eqnarray}
For the expansion to make sense $w$ should be sufficiently far ($\gg 1/j$)
away from $0$.  The coefficient $a_2$ will not be needed.
If we insert the expansions in the integral (\ref{Nint}) we encounter
functionals of the type
\begin{equation} \label{fun}
I[f]=\int d\eta\,\int d\nu f(\nu,\eta) e^{jS_0(\nu,\eta)}\,,
\end{equation} 
where $f(\nu,\eta)$ can be $1$ or any of the coefficients in (\ref{a0}) to
(\ref{b2}). The large parameter $j$ in the exponent suggests to integrate by
the two dimensional Laplace method.
To obtain correctly the first few orders in $1/j$ for $n(\tau)$ one
needs an extension of Laplace's method to higher
orders, i.e.~we also have to expand $I[f]$ in powers of $1/j$:
\begin{equation} \label{}
I[f]=\I{f}{0}+\frac{1}{j}\I{f}{1}+\frac{1}{j^2}\I{f}{2}+\ldots\,.
\end{equation}
The higher orders can be obtained from systematically 
reexpanding  in $1/j$ the following 
exact representation of $I[f]$ \cite{Fedoryuk87},
\begin{eqnarray}
I[f]&=&\re^{jS_0(\nu_0,\eta_0)}\frac{2\pi}{j}|\det(\bs)|^{-\frac{1}{2}}\sum_{l=0}^\infty\frac{1}{l!(2j)^l}(L_S)^l\left(f(\nu,\eta)\re^{jR(\nu,\eta)}
\right)\left.\right|_{\nu_0,\eta_0}\label{suml}\,,
\end{eqnarray}
where $(\nu_0,\eta_0)$ denotes the position
 of the maximum of $S_0$.
The matrix $\bs$ contains the four second
derivatives of $S$ at the  maximum, $\bs_{\nu,\eta}=\partial_\nu\partial_\eta
S_0\left.\right|_{\nu_0,\eta_0}$; the new ``action'' $R(\nu,\eta)$ is the
deviation of 
$S_0$ from its quadratic approximation, 
\begin{equation} \label{R}
R(\nu,\eta)\equiv S_0(\nu,\eta)-S_0(\nu_0,\eta_0)-\frac{1}{2}\langle
\bs {{\nu-\nu_0}\choose{\eta-\eta_0}},{{\nu-\nu_0}\choose{\eta-\eta_0}} 
\rangle\,,
\end{equation}
and $L_S$ is the second order homogeneous differential operator
\begin{equation} \label{LS}
L_S=\langle-\bs^{-1}{{\partial_\nu}\choose{\partial_\eta}},{{\partial_\nu}\choose{\partial_\eta}}
\rangle \,.
\end{equation}
The angular brackets denote a scalar product.
It is assumed that the integration range contains exactly one maximum with
$\partial_\nu S_0(\nu,\eta)=0=\partial_\eta S_0(\nu,\eta)$. Due to the
construction of $R$, all derivatives of $R$ of lower than third order vanish
and the term $L_S^l(f\re^{jR})$ in (\ref{suml}) is therefore a
polynomial in $1/j$ at most of the order $2l/3$. In order to get the first
order in $1/j$ beyond the usual Laplace method (the term with $l=0$) one has
therefore to go up to $l=3$.

Combining the expansions of $a$ and $b$ in terms of $a_k,b_k$ and the
expansions of the
corresponding functionals $I[a_k]$ and $I[b_k]$ according to (\ref{suml}) we
find
\begin{eqnarray}
n(\tau)&=&1+\ttau\Bigg[\frac{\I{a_0}{0 }}{\I{1}{0
}}-\frac{1}{j(\I{1}{0})^2}\big(\I{a_0}{0}\I 11-(\I{a_1}{0} \nonumber\\
&&+\I{a_0}{1})\I 10
\big)+{\cal O}\left(\frac{1}{j^2}\right)\Bigg]\nonumber\\
&&+\ttau^2\Bigg[\frac{\I{b_0}{0}}{\I 10}-\frac{1}{j(\I 10)^2}\big(\I{b_0}{0}\I 11-(\I{b_1}{0}+\I{b_0}{1})\I 10\big)\nonumber\\
&&-\frac{1}{j^2(\I{1}{0})^3}\Big( -(\I{b_2}{0}+\I{b_1}{1})(\I
10)^2+\big(\I{b_0}{0}\I 12 \nonumber\\
&&+\I 11\I{b_1}{0}+\I 11\I{b_0}{1}\big)\I 10-\I{b_0}{0}(\I
11)^2\Big)\Bigg]\,. \label{Irat}
\end{eqnarray}
The point of maximum $S_0$ is
located in our case  at 
\begin{equation} \label{nueta}
\eta_0=\frac{\gamma_2^2-\gamma_1^2}{(1+\gamma_1^2)(1+\gamma_2^2)}\mbox{, }\nu_0=\frac{1-\gamma_1^2\gamma_2^2}{(1+\gamma_1^2)(1+\gamma_2^2)}\,.
\end{equation}

To proceed further we distinguish two different cases.\\
{\bf Case 1: $\gamma_1\ne \gamma_2$} and {\bf $\gamma_1\gamma_2\ne 1$}\\
Here already the ordinary Laplace method  leads to a meaningful result,
\begin{eqnarray}
\frac{\I{a_0}{0}}{\I 10}&=&a_0(\nu_0,\eta_0)=-\frac{2(\gamma_1-\gamma_2)^2(1-\gamma_1\gamma_2)^2}{((1+\gamma_1^2)(1+\gamma_2^2))^2}\label{a0g}\\
\frac{\I{b_0}{0}}{\I 10}&=&b_0(\nu_0,\eta_0)=\frac{1}{2}a_0^2(\nu_0,\eta_0)\label{b0g}\,.
\end{eqnarray}
We have therefore
$n(\tau)=1+a_0(\nu_0,\eta_0)\ttau+\frac{1}{2}a_0^2(\nu_0,\eta_0)\ttau^2$,
and thus, correct up to ${\cal O}(\ttau^2)$ the anounced result (\ref{31}), 
where we have resubstituted $\ttau$ in terms of $j\tau$. \\

{\bf Case 2: $\gamma_1\gamma_2=1$}\\
The leading terms (\ref{a0g}), (\ref{b0g}) now vanish due to
$a_0(\nu_0,\eta_0)=0$. From the prefactor of the term linear
in $\ttau$ in (\ref{Irat}) only 
\begin{equation} \label{lin}
\frac{1}{j}\frac{\I{a_0}{1}}{\I 10}=-\frac{1}{j}\left(\frac{\gamma_1^2-1}{\gamma_1^2+1}\right)^2
\end{equation} survives, from
the quadratic term only 
\begin{equation} \label{qua}
\frac{1}{j^2}\frac{\I{b_2}{0}}{\I 10}=-\frac{1}{4j^2}(7\eta_0^2+1)\,,
\end{equation}
as the reader might verify in a straightforward but lengthy
calculation. In 
particular, the coefficient proportional to $1/j$ in the $\ttau^2$ term is
zero, such that to quadratic order in $\ttau$ $n(\tau)$ depends only on
$\tau$, not on $j\tau$:
\begin{equation} \label{ntauslow}
n(\tau)=1-\left(\frac{\gamma_1^2-1}{\gamma_1^2+1}\right)^2\tau-\frac{1}{4}\left(7\left(\frac{\gamma_2^2-\gamma_1^2}{(1+\gamma_1^2)(1+\gamma_2^2)}\right)^2+1\right)\tau^2+{\cal
O}(\ttau^3)
\end{equation}
The agreement of 
Eq.(\ref{ntauslow}) with exact numerical results extends well beyond the
range $\tau\ll 1/j$ for which the theory 
was made initially. This is not surprising since the main contribution
to the norm comes from a region where $n\simeq 0\simeq m$, so that the
limitation on the validity of the short time propagator is much less severe
than in the asumed worst case where both $n-m$ and $n+m$ are of order
$j$. The agreement with the numerical result becomes even better and leads
to rather  precise results even for $\tau\simeq 1$ if we  rewrite the decay
again in exponential form as was done in (\ref{32}).

\section{Possible experimental realization}
It has been experimentally verified that (\ref{eq:rhotd}) describes
adequately the radiation by identical atoms resonantly coupled to a leaky
resonator mode \cite{Haroche76} in a suitable parameter regime (see the
discussion in the context of the equation). It should therefore be possible
to observe 
the slow decoherence of the special Schr\"odinger cat states. We now propose a
scheme for their preparation.
Starting by all atoms in the ground state and with the field mode in its vacuum
state, a resonant laser 
pulse brings the Bloch vector to a coherent state $|\theta,\varphi\rangle$.
Note that the cavity may be strongly detuned (with respect to the atomic
transition frequency) during the whole preparation
process (detuning $\delta\gg \kappa$). The dissipation mechanism
(\ref{eq:rhotd}) is hereby
turned off and the system evolves unitarily with a Hamiltonian containing
a non--linear term $\propto (g^2/\delta) J_+J_-$. The free 
evolution during a suitable time will split the coherent state
$|\theta,\varphi\rangle$ in a superposition of $|\theta,\varphi'\rangle$ and
$|\theta,\varphi'+\pi\rangle$ as described in \cite{Agarwal97}. Finally, a 
resonant $\pi/2$ pulse brings the superposition to the desired orientation
symmetric to the equator, by rotation through the angle $\pi/2$ about an
axis perpendicular to the plane defined by the directions of the two
coherent states produced by the free evolution. At this point the
cavity can 
be tuned to resonance, thereby switching on the dissipation mechanism and
one can study the decay of coherence.\\

\section{Conlusion}

We have shown that a certain symmetry of the coupling to the environment
leads in the phenomen of superradiance to the existence of longlived
coherences of superpositions of macroscopically distinct quantum
states. Even though the components of the linear superposition are 
not exact eigenstates of the coupling operator to a degenerate eigenvalue,
the deviation from degeneracy is small enough for the coherences to decay on
a classical time scale only. We have proposed a preparation scheme with which
such Schr\"odinger cat states might be realizable experimentally.

{\em Acknowledgements:} We have enjoyed and profited from discussions with
Girish Agarwal, Serge   
Haroche, Walter Strunz, Dan Walls, and Peter Zoller. Financial support by
the Sonderforschungsbereich 
``Unordnung und gros\-se Fluktuationen'' of the Deutsche
Forschungsgemeinschaft is gratefully acknowledged.

\end{document}